# Smartphone-integrated RPA-CRISPR-Cas12a Detection System with Microneedle Sampling for Point-of-Care Diagnosis of Potato Late Blight in Early Stage

Jiangnan Zhao [a,b], Hanbo Xu [a,b], Cifu Xu [a,b], Wenlong Yin [a,b], Laixin Luo [c], Gang Liu [a,b], Yan Wang [a,b*]

[a] Key Laboratory of Smart Agriculture Systems, Ministry of Education, China Agricultural University, Beijing 100083, PR China

[b] Key Laboratory of Agricultural Information Acquisition Technology, Ministry of Agriculture and Rural Affairs of China, China Agricultural University, Beijing 100083, PR China

[c] Department of Plant Pathology, China Agricultural University, Beijing Key Laboratory of Seed Disease Testing and Control, Beijing, PR China.

**Author**

Jiangnan Zhao

College of Information and Electrical Engineering, China Agricultural University, China

E-mail address: zhaojiangnan@cau.edu.cn

Hanbo Xu

College of Information and Electrical Engineering, China Agricultural University, China

E-mail address: 2022308130516@cau.edu.cn

Cifu Xu

College of Information and Electrical Engineering, China Agricultural University, China

E-mail address: xucifu@cau.edu.cn

Wenlong Yin

College of Information and Electrical Engineering, China Agricultural University, China

E-mail address: 360224638@qq.com

Prof. Laixin Luo

Department of Plant Pathology, China Agricultural University, China

E-mail address: luolaixin@cau.edu.cn

Prof. Gang Liu

College of Information and Electrical Engineering, China Agricultural University, China

E-mail address: pac@cau.edu.cn

**\* Corresponding author**

A/Prof. Yan Wang

Mailing address: College of Information and Electrical Engineering, China Agricultural University, China.

E-mail address: yanwang@cau.edu.cn

Tel: +86-18501182599

# Abstract

Potato late blight, caused by the oomycete pathogen *Phytophthora infestans*, is one of the most devastating diseases affecting potato crops in the history. Although conventional detection methods of plant diseases such as PCR and LAMP are highly sensitive and specific, they rely on bulky and expensive laboratory equipment and involve complex operations, making them impracticable for point-of care diagnosis in the field. Here in this study, we report a portable RPA-CRISPR based diagnosis system for plant disease, integrating smartphone for acquisition and analysis of fluorescent images. A polyvinyl alcohol (PVA) microneedle patch was employed for sample extraction on the plant leaves within one minute, the DNA extraction efficiency achieved 56 μg/mg, which is ~3 times to the traditional CTAB methods (18 μg/mg). The system of RPA-CRISPR-Cas12a isothermal assay was established to specifically target *P. infestans* with no cross-reactivity observed against closely-related species (*P. sojae*, *P. capsici*). The system demonstrated a detection limit of 2 pg/μL for *P. infestans* genomic DNA, offering sensitivity comparable to that of benchtop laboratory equipment. The system demonstrates the early-stage diagnosis capability by achieving a ~80% and 100% detection rate on the third and fourth day post-inoculation respectively, before visible symptoms observed on the leaves. The smartphone-based "sample-to-result" system decouples the limitations of traditional methods that rely heavily on specialized equipment, offering a promising way for early-stage plant disease detection and control in the field.

# 1. Introduction

Late blight of potato, caused by the oomycete *Phytophthora infestans*, is the most destructive disease affecting potato crops (Fry et al., 2015; Kong et al., 2020). Historically, it was responsible for the Irish Great Famine in the 1840s, which resulted in the deaths or displacement of millions and caused a dramatic reduction in Ireland's population by nearly one quarter (Anderson et al., 2004; Haas et al., 2009; Tang et al., 2023). *P. infestans* infects various parts of the potato plant, including leaves, tubers, and stems, and is often referred to as the "cancer of potatoes." The disease spreads rapidly and extensively; once symptoms appear, it can devastate entire potato fields within two to three weeks under cool and moist conditions (Austin Bourke, 1964; Fry, 2008; Ristaino et al., 2020). Therefore, rapid detection and diagnosis are crucial for effective management of late blight, particularly during the early stages of disease development.

Currently, the detection of *P.infestans* primarily relies on molecular techniques such as polymerase chain reaction (PCR) and loop-mediated isothermal amplification (LAMP). As early as 1997, Trout et al. developed a PCR-based method targeting the ITS gene for the detection of *P.infestans* and its closely related species *Phytophthora mirabilis* and *Phytophthora cactorum*. However, PCR-based detection requires precise thermal cyclers and controlled laboratory conditions, along with skilled personnel, which limits its applicability for rapid on-site detection (Hieno et al., 2019; Ristaino et al., 2020; Zhen et al., 2016). In 2016, Hansen et al. designed the first colorimetric LAMP assay targeting the ITS gene for rapid detection of *P.infestans* DNA. Khan et al. (2017) further reported a LAMP assay targeting the Ypt1 gene, which encodes a Ras-related protein, suitable for *P.infestans* detection. Additionally, Kong et al. (2020) developed the LAMP-PiSMC system, which targets the species-specific PiSMC nucleotide sequence for early diagnosis of *P.infestans*. Despite their advantages, LAMP assays typically require relatively high isothermal temperatures (60~65 °C) and are susceptible to primer-dimer formation (Zhao et al., 2023), posing challenges for their application in field-based rapid diagnostics.

The clustered regularly interspaced short palindromic repeats and associated protein system (CRISPR-Cas) is a natural immune mechanism found in prokaryotes (Weng et al., 2023). Due to its high sensitivity, specificity, and operational simplicity, the CRISPR-Cas system has shown great potential in gene editing and molecular diagnostics (Jiang et al., 2023; G. Liu et al., 2022). Upon binding to its specific target, the system exhibits nonspecific trans-cleavage activity, enabling efficient and specific detection of target pathogens (Bai et al., 2019; Li et al., 2018; Liu et al., 2021). To enhance the analytical performance of CRISPR-Cas-based detection, it is often coupled with amplification techniques such as recombinase polymerase amplification (RPA) (Wang et al., 2019, 2020), loop-mediated isothermal amplification (LAMP) (Zhang et al., 2021), and PCR (Zhou et al., 2020) to amplify target genes. Depending on the type of nucleic acid substrate (DNA or RNA), CRISPR-Cas systems have been applied for detecting a wide range of pathogens, including viruses (Fozouni et al., 2021; Kellner et al., 2019), bacteria (L. Liu et al., 2022; Wang et al., 2020; Xu et al., 2022), and parasites (Li et al., 2022; Ma et al., 2021). However, research on the application of CRISPR-based detection in oomycetes remains lacking, and a CRISPR system specifically designed for the detection of *P.infestans* has yet to be established.

Based on this, the present study is the first to apply CRISPR-Cas12a technology for the detection of the potato late blight pathogen. We developed an RPA-CRISPR-Cas12a assay targeting *P.infestans* genomic DNA, which enables specific detection with a sensitivity as low as 2 pg/μL. Inspired by the study of Paul et al. (2020), polyvinyl alcohol (PVA) was used to fabricate microneedle patches for genomic DNA extraction. This approach significantly reduced the conventional DNA preparation time from 3–4 hours to just 1 minute. The entire workflow from DNA extraction to result visualization can be completed within 90 minutes. To facilitate use in non-specialized laboratory environments and field settings, we also developed a portable smartphone-based fluorescence detection device. Combined with microneedle DNA sampling and the RPA-CRISPR-Cas12a system, this platform enables rapid, highly sensitive, and field-deployable detection of *P.infestans*, providing a valuable tool for early warning and management of potato late blight.

The complete detection workflow is illustrated in **Fig.1**. A single potato leaf is sampled, and a microneedle patch is pressed onto the leaf surface for 60 seconds to extract DNA. After removal, the microneedle patch is transferred into a centrifuge tube containing the RPA reaction mixture and incubated at 37 °C for 20 minutes. Subsequently, 2 μL of the RPA product is added to a preassembled CRISPR-Cas12a detection system and incubated at 37 °C for 60 minutes within a portable fluorescence detection device. During the detection process, fluorescence images are automatically captured every 5 minutes using the smartphone camera. Upon completion, a dedicated application analyses the image data and generates a real-time fluorescence intensity curve. This integrated detection system enables fully automated operation from sample preparation to result interpretation within 90 minutes. It is characterized by ease of use and intuitive output, making it particularly suitable for on-site and real-time diagnosis of potato late blight under field conditions.

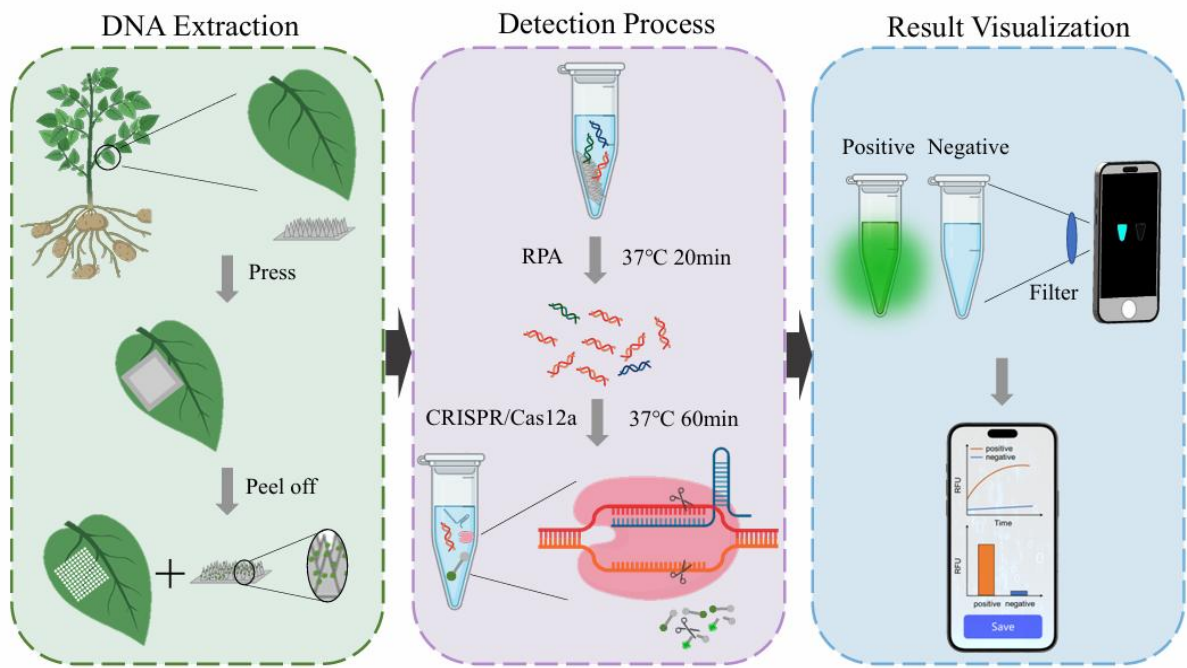

**Fig. 1.** Schematic diagram of the application of the MN-RPA-Cas12a detection system on potato late blight.

## 2. Materials and methods

*2.1 Materials*

All primers and crRNAs used in this study were synthesized by Sangon Biotech (Shanghai, China). The LbCas12a protein and reaction buffer were also purchased from Sangon Biotech. The RPA assay kit was obtained from TwistDx Ltd. (Cambridge, Massachusetts, USA). DNA oligonucleotides were synthesized by Shanghai GeneBiogist Technology Co., Ltd. (Shanghai, China). Polydimethylsiloxane (PDMS) mold (13.3 mm × 13.3 mm in size, 15 × 15 microneedle conical cavity array, 840 μm in height, and 280 μm base diameter) was provided by Taizhou Microchip Medical Technology (Taizhou, China). The strains used in this study, including *Phytophthora infestans*, *Phytophthora sojae*, and *Phytophthora capsici*, were provided by the Laboratory of Fungicide Pharmacology & Pathogen Resistance, China Agricultural University. Healthy potato leaves were collected from Rizhao City, Shandong Province, China. A field survey was conducted prior to sampling to confirm the absence of potato late blight in the field.

*2.2 Target selection and system design*

Based on the study by Tang et al. (2023) , a 1050 bp species-specific genomic fragment of *P.infestans*, located at 351~1400 of the scaffold NW_003302563.1, was selected. This sequence was cloned into the pUC57 (Sangon Biotech, Shanghai, China) to generate a reference plasmid, designated PiNW. Three forward primers and four reverse primers, each 30 nt in length, were designed within this region. The design was carried out using DNAMAN software (Version 9.0, Lynnon Co., USA) and followed the TwistDx® user manual. Primer sequences are listed in **Table 1**. A 23 bp region within the RPA amplification product was selected as the CRISPR-Cas12a target site. This target sequence, containing a PAM site (TTTN-), was used for crRNA design. In **Fig. 2a**, the blue region represents the target sequence, while the red region corresponds to the complementary crRNA. All primers, crRNAs, and ssDNA used in the experiments were stored at –80 °C until use.

**Table 1.** Primers, Probe, and crRNA used in the detection system.

| Name | Sequence | Length(nt) |
| --- | --- | --- |
| RPA-F1 | AACTAACGGTTCTCCTACTTCAACAGTGGG | 30 |
| RPA-F2 | CTAACGGTTCTCCTACTTCAACAGTGGGAC | 30 |
| RPA-F3 | GATACTATTGCAGGCTGGATACTGTAGATG | 30 |
| RPA-R1 | TTTCTGAGCGTAGGCACTTTATTTGTCTTC | 30 |
| RPA-R2 | CTGTAGGTCCATTCCGTAGACAAGACGAGC | 30 |
| RPA-R3 | TCTGTAGGTCCATTCCGTAGACAAGACGAG | 30 |
| RPA-R4 | TTTCTGTAGGTCCATTCCGTAGACAAGACG | 30 |
| crRNA | UAAUUUCUACUAAGUGUAGAUAAGCGAUCGUUCAAAAUUUUACC | 44 |
| FQ reporter | FAM-TTATTATT-BHQ1 | 8 |

## 2.3 Microneedle patch fabrication

The fabrication method of the microneedle patch was adapted from the report by Paul et al. (2021). Before fabricating the microneedle (MN) patch, the MN mold was placed in an ultrasonic bath for 5 minutes for cleaning. Then, the mold was wiped with a tissue to remove excess surface moisture. Subsequently, 500 μL of 15% (w/v) PVA solution was added to the mold. To facilitate the infiltration of the PVA solution into the needle cavities, the mold was placed in a vacuum chamber containing silica gel and subjected to vacuum for 20 minutes. After turning off the vacuum pump, the mold was kept in the sealed chamber for an additional 15~20 minutes to allow air bubbles to dissipate. Once removed from the vacuum pump, the mold was left at 25°C for 24 hours to ensure the complete drying of the PVA solution. After drying, the MN patch was carefully peeled from the mold. A well-formed patch displayed a uniform 15 × 15 conical microneedle array, with each needle having a tip height of 840 μm and a base diameter of 280 μm (**Fig. 4a, b**). The fabricated patches were stored at room temperature in sealed Petri dishes for subsequent use.

## 2.4 DNA extraction

This study utilized two DNA extraction methods: a cetyltrimethylammonium bromide (CTAB)-based protocol and a microneedle patch-based approach.

### 2.4.1 CTAB-Based DNA Extraction

Genomic DNA was extracted from *P.infestans* mycelium and potato leaves following the CTAB method (Zhang S, 2023) with minor modifications. Approximately 0.5 g of potato leaf tissue or fungal mycelium was homogenized in a 2 mL centrifuge tube containing 800 μL of pre-heated extraction buffer (LS0006, Solarbio®, Beijing, China) at 65°C, along with two 5 mm grinding beads, and vortexed for 30 min. The homogenate was then incubated at 65°C in a dry heat block for 1 hour, with brief vortexing at 10-minute intervals. Following a 2-minute cooling period, 600 μL of phenol:chloroform:isoamyl alcohol (25:24:1) was added to each sample. The mixture was agitated vigorously for 3 minutes to ensure complete emulsification before centrifugation at 12,000 rpm for 10 minutes at room temperature. The aqueous phase (500 μL) was carefully transferred to a fresh 1.5 mL tube, avoiding contamination from the interphase or organic layer. DNA was precipitated by adding an equal volume of isopropanol, gently inverting the tube, and incubating at −20°C for 30 minutes. Following centrifugation at 12,000 rpm for 10 minutes, the supernatant was discarded, and the DNA pellet was washed with 1 mL of 75% ethanol (without vortexing) and centrifuged again at 12,000 rpm for 5 minutes. After air-drying the pellet for 15 minutes at room temperature, the DNA was resuspended in 50 μL of nuclease-free water or TE buffer(pH 8.0) and stored at −20°C until further use.

*2.4.2 Microneedle Patch-Based DNA Extraction*

To streamline DNA extraction, a microneedle patch-based method (Paul et al., 2021) was employed for both potato leaves and *P. infestans*. Briefly, a sterile microneedle patch was applied to the infected leaf region and pressed gently for 1 minute. The patch was then removed, rinsed with 50 μL of TE buffer (pH 8.0), and the eluate was collected and stored at −20°C for downstream applications.

## 2.5 Establishment of RPA-CRISPR-Cas12a detection

The standard RPA reactions were performed according to the instructions of the TwistAmp Basic kit. Each reaction contained one pellet, 29.5 μL of rehydration buffer, 480 nM each of forward and reverse primers, 2 μL of the DNA template, and 14 mM of magnesium acetate (MgOAc). The final reaction volume was adjusted to 50 μL with nuclease-free water. The reaction mixture was incubated at 37°C for 4 minutes, inverted 10 times, and briefly centrifuged. The amplification was then continued at 37°C for an additional 16 minutes.

The Cas12a-mediated trans-cleavage fluorescence detection consisted of 100 nM LbCas12a, 100 nM crRNA, 200 nM ssDNA-FQ reporter, 5 μL of reaction buffer, and 2 μL of RPA-amplified product, with sterile water added to bring the final reaction volume to 50 μL. The ssDNA-FQ reporter was a non-specific single-stranded DNA probe that was labeled with a 6-FAM fluorophore at the 5′ end, , which has an excitation wavelength of 495 nm and an emission wavelength of 521 nm, and a BHQ1 quencher at the 3′ end. The reactions were carried out in a 96-well plate and incubated at 37 °C for 60 minutes using a QuantStudio 3

Real-Time PCR System (Thermo Fisher Scientific, Sunnyvale, CA, USA), with fluorescence measurements recorded every 5 minutes.

*2.6 Real-Time PCR Detection*

The standard qPCR reactions were performed according to the instructions of the 2×SG Fast qPCR Master Mix (Low Rox) (Sangon Biotech, Shanghai, China). Each reaction contained 10 μL of 2× master mix, 200 nM each of forward and reverse primers, an appropriate amount of genomic DNA template, and nuclease-free water to adjust the final reaction volume to 20 μL. The PCR cycling conditions were set according to the instructions: an initial denaturation at 95 °C for 3 minutes, followed by 40 cycles of denaturation at 95 °C for 3 seconds, and annealing and extension at 60 °C for 30 seconds.

*2.7 Design of the smartphone-based detection device*

We designed and developed a low-cost, portable, and integrated detection device that highly integrates RPA isothermal amplification, CRISPR-Cas12a reaction, and fluorescence signal detection, combined with a smartphone-based real-time analysis system to achieve rapid field detection from "sample-in to result-out" (**Fig. 7a, b, c**). The device housing was fabricated using light-cured resin through 3D printing technology (dimensions: 20 cm × 12 cm × 17.4 cm), with a total weight of only 450 g. The internal structure integrates the reaction zone, optical system, and temperature control unit.

The reaction zone accommodates four centrifuge tubes, allowing simultaneous operation of four RPA-CRISPR-Cas12a reactions. Each tube has a volume of 1.5 mL, sufficient to hold a 50 μL reaction mixture. Rectangular partitions (height: 3.8 cm; width: 1.5 cm) are installed between adjacent reaction tubes to prevent cross-interference of fluorescence signals. To heat the RPA-CRISPR-Cas12a reaction mixture, a resistance wire heating pad was mounted on the rear wall of the reaction zone. This heating element supports stable incubation at 37 °C, and it is connected to a temperature controller that allows real-time monitoring and setting of reaction temperatures. A laser with an excitation wavelength of 470 nm was used. The laser is transmitted through optical fibers (1.0 mm diameter) to the bottom of each reaction tube, vertically illuminating the tube to excite the FAM fluorophores. Fluorescence signals are directly captured by the smartphone camera. A bandpass filter (520 ± 15 nm) is placed in front of the camera lens to block stray light and separate the fluorescence signal from the excitation background, ensuring a high signal-to-noise ratio. Both the heating pad and the laser are powered by a portable power bank, which, when fully charged, can operate the entire device for up to three hours, sufficient to complete the detection workflow. The total cost of the device, excluding the smartphone, is within 800 RMB, making it suitable for plant pathogen monitoring in resource-limited settings.

The device is very easy to use: it is placed on a flat surface, and the power bank is turned on to preheat the device to 37 °C. After preparing the reaction mixtures and loading them into the reaction zone, the detection can begin. Simultaneously, the custom-developed smartphone application is launched, which automatically captures one image of the reaction system every five minutes. After the reaction is complete, the application analyses the fluorescence intensity change curves for each sample, facilitating rapid interpretation of the detection results.

## 3. Results and discussion

### *3.1 RPA primer screening*

To ensure optimal detection sensitivity of the RPA-CRISPR-Cas12a assay, cross-reaction screening of RPA primers was performed using the RPA reaction system described in Section 2.4. The RPA amplification products were analysed by 2% agarose gel electrophoresis. The results showed that amplification products were clearly observable for all primers, with RPA-F3 and RPA-R1 displaying particularly strong target bands and the highest amplification efficiency (**Fig. 2b, c**).

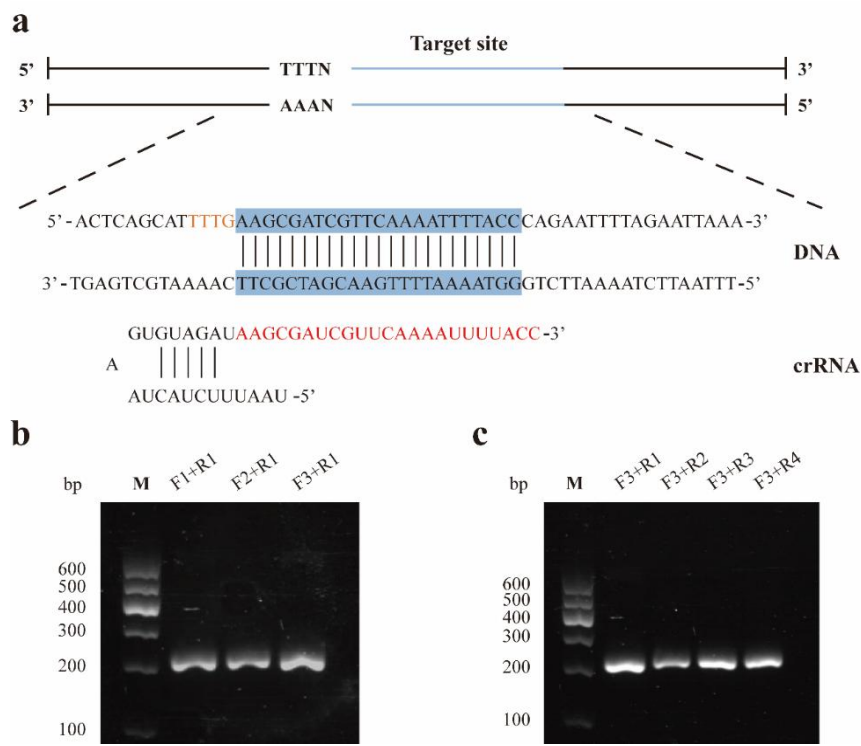

**Fig. 2**. **(a)** Schematic of dsDNA target detected with the Cas12a/crRNA. crRNA was provided in detail. The dsDNA target site was highlighted with blue shading. and crRNA fragments in red. **(b)** Screening for the best RPA forward primers. **(c)** Screening of RPA

reverse primers using preferred forward primers. All reactions were repeated three times. M: DNA marker

### *3.2 Sensitivity and specificity of RPA-CRISPR-Cas12a*

We explored the detection sensitivity of the RPA-CRISPR-Cas12a assay for the recombinant plasmid PiNW and genomic DNA. The plasmid PiNW (10ng/μL) was serial diluted (10-fold) and 2 μL plasmid of each dilution was used as the template for RPA amplification. Final plasmid concentrations ranging from $10^4$ to $10^{-2}$ pg/μL were tested. Reactions were performed at 37 ℃ for 60 minutes, with fluorescence measurements taken every 5 minutes. Each concentration was tested in triplicate. The endpoint fluorescence results showed that fluorescence intensity decreased with decreasing plasmid concentration. When the plasmid concentration reached $10^{-2}$ pg/μL, the fluorescence intensity was no longer significantly different from that of the negative no-template control. As shown in **Fig. 3a, c** the fluorescence detection sensitivity for plasmid PiNW was determined to be $1\times10^{-1}$ pg/μL. A linear regression analysis was performed between the logarithm of plasmid concentration and the relative fluorescence units (**Fig. 3e**), yielding an $R^2$ value of 0.8991.

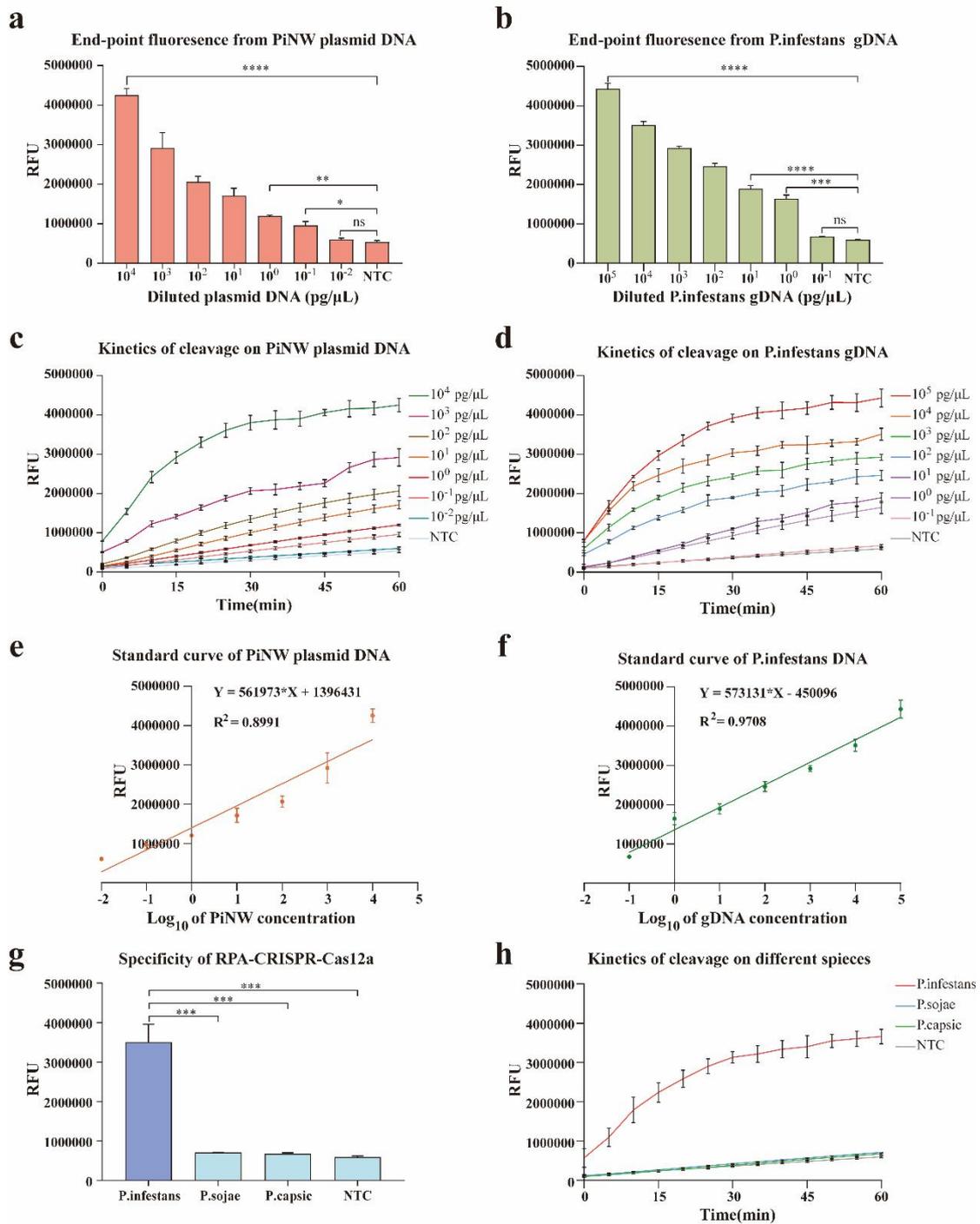

**Fig. 3.** Establishment of RPA-CRISPR-Cas12a detection. **(a, c, e)** Sensitivity analysis of RPA-CRISPR-Cas12a detection on plasmid PiNW. **(b, d, f)** Sensitivity analysis of RPA-CRISPR-Cas12a detection on *P.infestans* gDNA. **(g, h)** RPA-CRISPR-Cas12a assay specificity analysis among three different fungus, namely, *P.infestans*, *P.sojae*, and *P.capsic*. n

= 3 biological replicates, two-tailed Student's t test; * $P < 0.05$, ** $P < 0.01$, *** $P < 0.001$, and **** $P < 0.0001$; *n.s.* not significant; NTC nontarget control; bars represent mean ± S.D.

To determine the detection sensitivity of the RPA-CRISPR-Cas12a assay for *P.infestans* gDNA, DNA was extracted from *P. infestans* mycelium using the CTAB method. The extracted gDNA (20 μg/μL) was serially diluted (10-fold) with sterile water to obtain final concentrations ranging from $10^5$ to $10^{-1}$ pg/μL. The end-point fluorescence and kinetic results showed that the RPA-CRISPR-Cas12a assay had a detection sensitivity of approximately 2pg/μL(**Fig. 3a, c**) for *P.infestans* gDNA. The standard curve obtained by linear regression (**Fig. 3f**) showed an $R^2$ value of 0.9708. This result is comparable to the detection limit achieved by Paul et al. using the real-time LAMP method.

We selected *Phytophthora infestans*, *Phytophthora sojae*, and *Phytophthora capsici* to evaluate the specificity of the RPA-CRISPR-Cas12a assay. Genomic DNA was extracted from the mycelia of *P. infestans*, *P. sojae*, and *P. capsici* using the CTAB method, and ddH$_2$O was used as a negative control. As shown in **Fig. 3g, h**, only the gDNA isolated from *P. infestans* was able to activate Cas12a and produce a detectable fluorescence signal, indicating that the RPA-CRISPR-Cas12a assay exhibited high specificity without cross-reactivity to other pathogens. This specificity was provided by both the RPA primers and the crRNA.

### 3.3 Microneedle-Based DNA Extraction Combined with RPA-CRISPR-Cas12a Assay

Although the CTAB method is currently the mainstream and widely used approach for genomic DNA extraction, it involves multiple tedious experimental steps such as cell lysis, precipitation, washing, and dissolution. On average, DNA extraction using the CTAB method takes approximately 3 hours, which often becomes a major obstacle for timely on-site detection. In 2019, Paul et al. proposed a method for extracting DNA using microneedle patches made of PVA. PVA is a hydrophilic polymer material that, when inserted into plant leaves, can disrupt plant cells and simultaneously absorb intracellular nucleic acids. This method enables DNA extraction within just 1 minute, significantly reducing the time required for sample preparation.

To compare the DNA extraction efficiency between the MN patch method and the CTAB method, DNA was isolated from *P.infestans*-infected potato leaves (6 days post-inoculation) using both approaches. First, DNA was extracted using the microneedle patch as described in Section 2.4.2. Plant tissue fragments from the potato leaves were visibly attached to the microneedle patch after puncturing **(Fig. 4c, d)**. Subsequently, the corresponding leaf area punctured by the microneedle was excised and subjected to DNA extraction using the CTAB method. The DNA concentrations obtained by both methods were measured using a NanoDrop One Microvolume UV−vis spectrophotometer. Meanwhile, the weight change of the microneedle patch before and after puncture, as well as the mass of leaf tissue processed by the CTAB method, were recorded to generate **Fig. 4e.**

The experimental results showed that 50 ng of DNA was extracted using the MN patch, while 235 ng of DNA was extracted using the CTAB method. Although the total DNA yield was lower for the microneedle patch, this was attributed to the significantly smaller sampling mass—only 0.9 mg for the MN method compared to 14 mg for the CTAB method.

The average DNA yield per milligram of leaf tissue was 56 µg for the MN method and 18 µg for the CTAB method **(Fig. 4f)**, indicating that the microneedle patch exhibited a much higher DNA extraction efficiency.

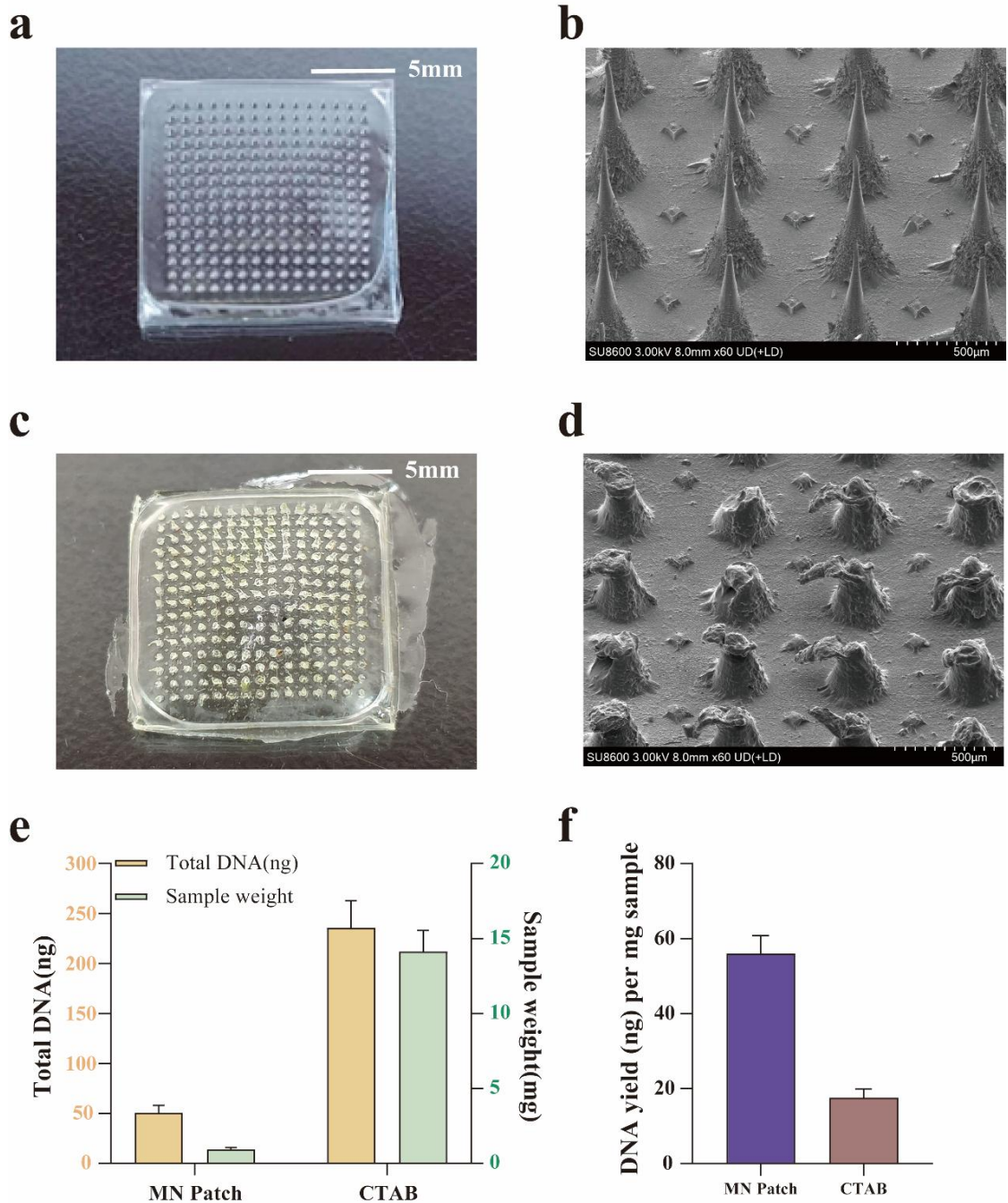

**Fig. 4.** Comparative analysis of MN Patch and CTAB protocols for plant DNA extraction. **(a,**

**b)** A photograph and Scanning electron microscopy image of a fresh MN patch, respectively. **(c, d)** A photograph and Scanning electron microscopy image of a MN patch after leaf puncture. Scale bar: 5 mm for **(a, c)** and 500 μm for **(b, d)**. **(e)** Total DNA amounts extracted by the two methods (yellow bars). Differences in sampling volume between the two methods (green bars). **(f)** Comparison of the DNA yields (N = 5) obtained from the MN patch and conventional CTAB RNA extraction protocol.

Subsequently, to verify that DNA extracted by the microneedle (MN) patch could be directly used for molecular detection, the unpurified and undiluted MN-extracted DNA was subjected to qPCR and RPA-CRISPR-Cas12a assays. In the qPCR assay, previously published primers targeting a 100 bp region of the *P. infestans* ITS gene (Paul et al., 2020) were used. DNA extracted by the CTAB method served as the positive control, while DNA from healthy potato leaves without *P. infestans* infection was used as the negative control. As shown by the real-time amplification curves **(Fig. 5a)**, *P. infestans* was successfully detected in both the MN- and CTAB-extracted samples. However, due to the lower DNA concentration in the MN-extracted sample, the Ct value (11.8) was higher compared to that obtained by the CTAB method (8.7) (Fig. 5b).

In the RPA-CRISPR-Cas12a assay, the kinetic fluorescence curves **(Fig. 5c)** and end-point fluorescence intensity values **(Fig. 5d)** demonstrated that the fluorescence signal from the MN-extracted sample was significantly higher than that of the negative control group, indicating that *P. infestans* was accurately detected using this approach. These results confirm that DNA extracted via MN patches is suitable for molecular detection, greatly simplifying the conventional DNA extraction process. Although the amount of DNA extracted from plant leaves using the MN patch was relatively low, it was still sufficient for downstream detection. Furthermore, the MN-extracted DNA could be directly used for qPCR and RPA-CRISPR-Cas12a detection without the need for additional purification.

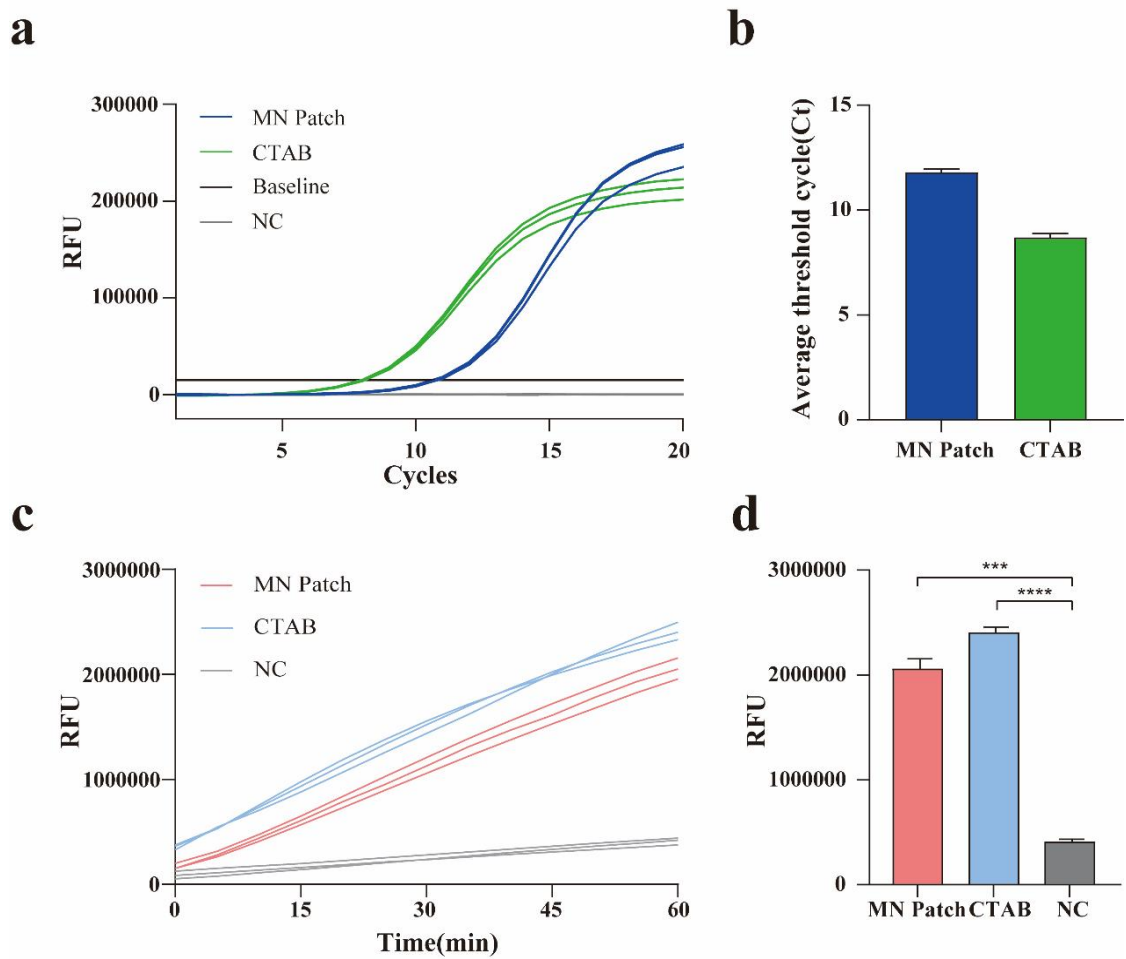

**Fig. 5.** MN patch-based DNA extraction directly applicable to molecular assays without further purification. **(a, b)** Real-time RT-PCR amplification of *P.infestans* DNA extracted by the MN patch and CTAB protocol. **(c, f)** RPA-CRISPR-Cas12a detection of *P.infestans* DNA extracted by the MN patch and CTAB protocol.

*3.4 Early Detection of Potato Late Blight*

Using the established microneedle-based DNA extraction combined with the RPA-CRISPR-Cas12a assay, further investigation was conducted for the detection of potato late blight. Potato leaves with uniform size and healthy growth were selected. The leaves were first rinsed with tap water to remove surface dust, then disinfected by washing with 75% ethanol for 2 seconds, followed by three rinses with sterile water (Zhang W, 2024). The leaves were placed abaxial side up on filter papers moistened with sterile water in Petri dishes. Each leaf was evenly sprayed with 500 μL of *P. infestans* spore suspension at a concentration of $4 \times 10^4$ spores/mL. The Petri dishes were incubated in the dark at 18 °C for 24 hours.

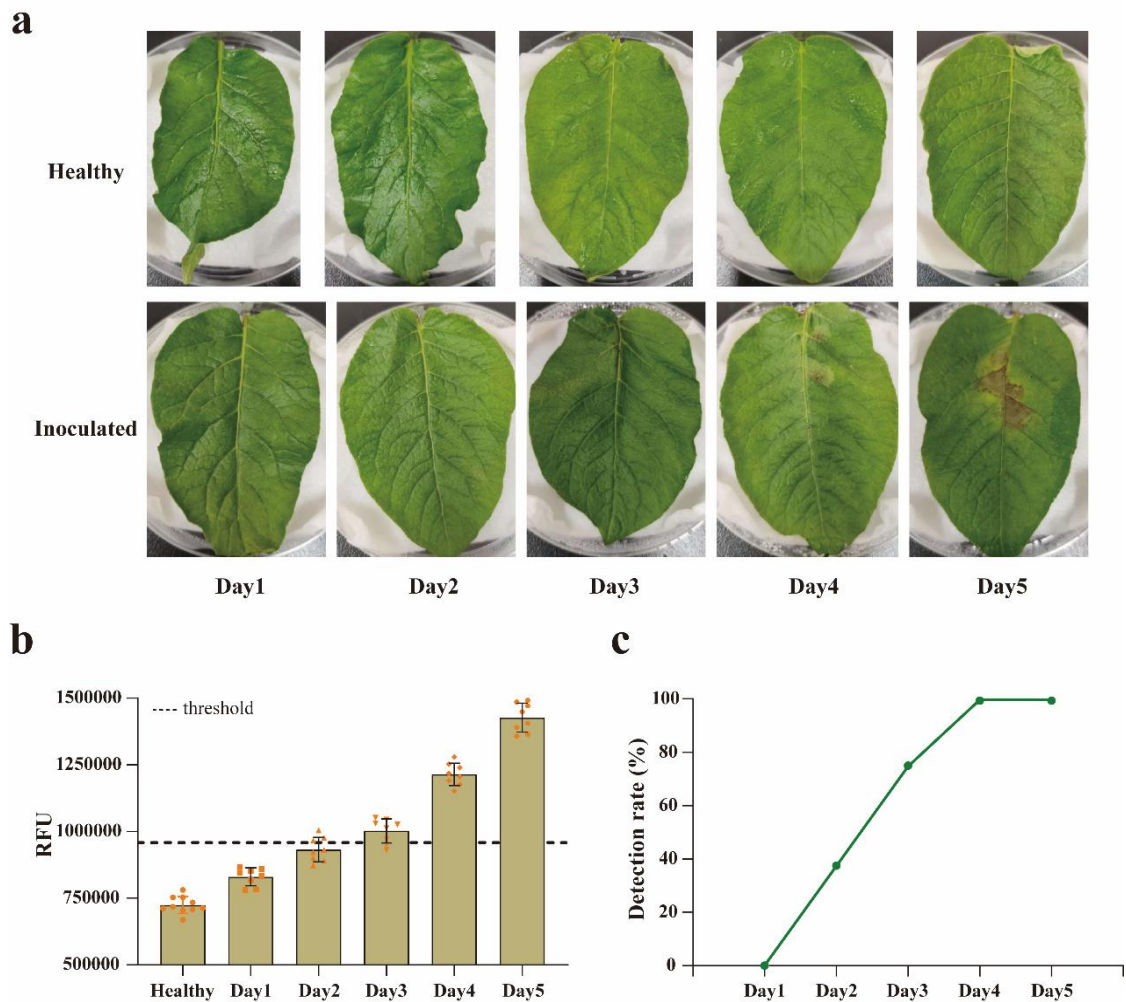

**Fig. 6.** Application of MN extraction for RPA-CRISPR-Cas12a detection of *P. infestans* from laboratory-inoculated samples. **(a)** Temporal changes in healthy uninoculated (top) and inoculated (bottom) leaves over five days post-inoculation (days 1−5). **(b)** Comparison of MN-RPA-Cas12a detection results across different inoculation days. **(c)** Detection sensitivities for different inoculation days.

After incubation, residual droplets of the spore suspension were wiped off, and the leaves were flipped with the adaxial side facing up. They were then transferred to an environment of 22 °C under a 16-hour light/8-hour dark cycle for further incubation (Yang J, 2023) . A total of 20 leaves were inoculated, with an additional 5 non-inoculated healthy leaves serving as negative controls. From day 1 to day 5 post-infection, at the same time each day, DNA was extracted using the MN patch from two different positions on each of four infected leaves and one healthy control leaf.

**Fig. 6a** shows the growth status of healthy and infected leaves at different time points post-inoculation. During the 5 days following inoculation, the healthy leaves exhibited no significant changes in appearance, whereas the infected leaves began to develop late blight lesions on day 4. By day 5, the lesions had expanded further and became prominent. The DNA extracted from the leaves using the MN patches was subjected to RPA-CRISPR-Cas12a detection, and the end-point fluorescence intensities are shown in **Fig. 6b**. In the experimental group, fluorescence intensity increased progressively with the duration of infection, indicating a gradual accumulation of *P. infestans* within the leaves.

To distinguish positive samples from negative ones during detection, a threshold value was set: samples with fluorescence intensity above the threshold were considered positive, whereas those below were considered negative. The principle of threshold setting was to minimize false-negative and false-positive results, which is critical for ensuring detection accuracy. The threshold was calculated as the mean fluorescence intensity of the negative samples plus three times the standard deviation (Mean + 3×SD), based on the assumption of a normal distribution, covering 99.7% of negative samples with a false-positive rate below 0.15%. In this study, the sterile water was used as the negative control and tested 30 times, yielding a mean fluorescence intensity of 720,444.91 and a standard deviation (SD) of 79,298.87. Thus, the threshold was determined to be 958,341.52.

As shown in **Fig. 6b**, on the second day post-inoculation, 3 out of 8 experimental samples exceeded the threshold; on the third day, 6 samples exceeded the threshold; and by the fourth day, all samples showed fluorescence intensities above the threshold, indicating that all infected leaves could be detected. The detection rate was defined as the number of true positives divided by the total number of positives. As shown in **Fig. 6c**, the detection rate reached 100% on day 4 post-inoculation, even though visible symptoms on the leaves were still minimal (**Fig. 6a**). These results demonstrate that the microneedle-based DNA extraction combined with the RPA-CRISPR-Cas12a assay enables the early detection of potato late blight infection with high sensitivity.

### *3.5 Smartphone-coupled MN-RPA-CRISPR-Cas12a detection*

We investigated the detection sensitivity of the smartphone-based detection device for *P. infestans* DNA. Genomic DNA was extracted from *P. infestans* mycelium using the CTAB method (40 μg/μL) and serially diluted with sterile water to final concentrations ranging from $10^4$ to $10^{-1}$ pg/μL across six gradients. Each diluted sample was prepared into the corresponding RPA-CRISPR-Cas12a reaction mixture and placed into the smartphone-based detection device for reaction.

The normalized fluorescence intensity for each concentration was calculated by dividing the change in fluorescence intensity during the reaction by the initial fluorescence intensity, as described by Paul et al. (2021). As shown in **Fig. 7d–f**, the experimental results demonstrated that the smartphone-based detection device exhibited fluorescence enhancement over time similar to that observed using a benchtop PCR machine. Moreover, a linear relationship was observed between the endpoint fluorescence intensity and the logarithm of the target DNA concentration. The detection sensitivity achieved by the smartphone device was as low as 4 pg/μL, which was comparable to that of the benchtop PCR system.

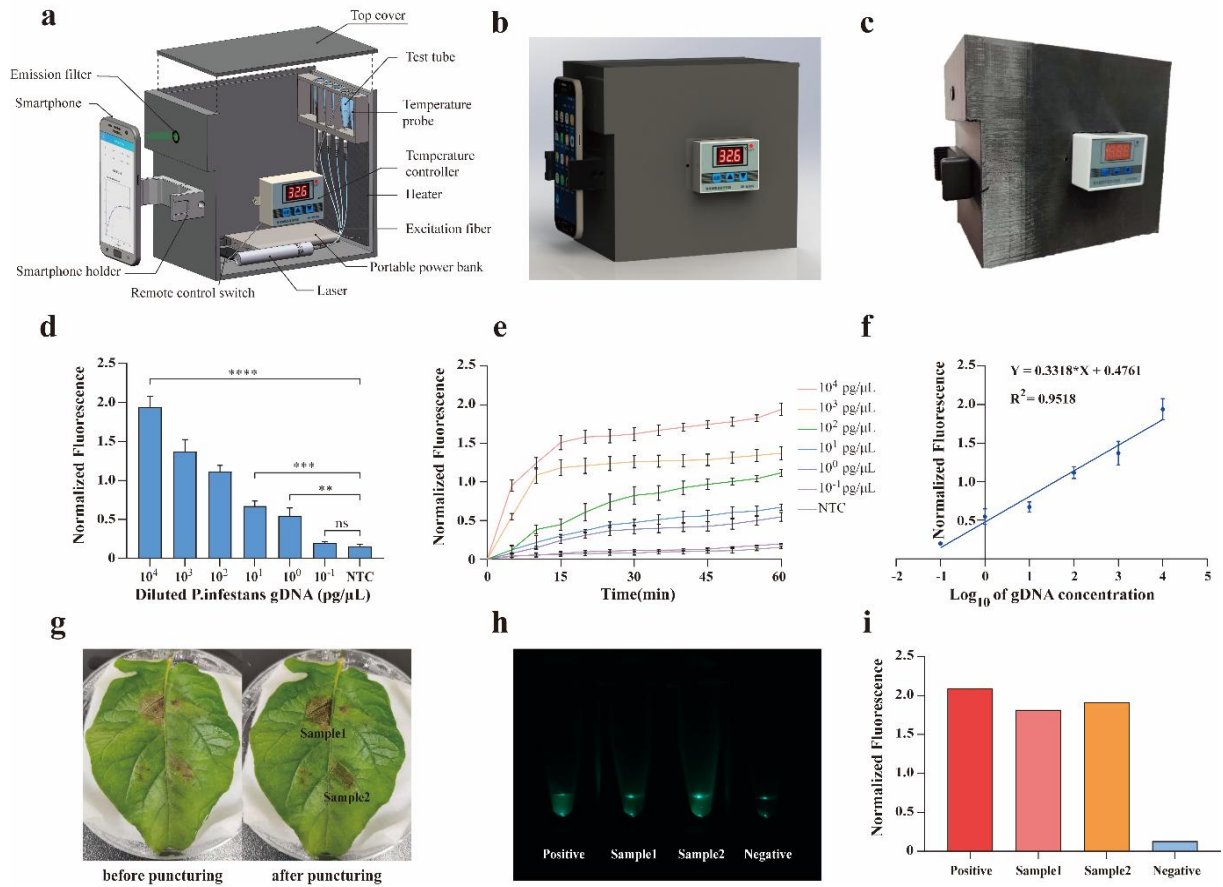

**Fig. 7.** Smartphone-integrated RPA-CRISPR-Cas12a detection system with microneedle sampling. **(a)** Cross-section view of smartphone reader device showing its internal components. **(b, c)** A design diagram and photograph of the smartphone reader. **(d, e, f)** Sensitivity analysis of the smartphone-based RPA-CRISPR-Cas12a detection platform. All error bars indicate SD for N = 6 samples (two separate runs with 3 samples in each run). **(g, h, i)** Detection of *P.infestans* in potato leaves using the integrated MN-smartphone RPA-CRISPR-Cas12a platform. **(g)** Images of late blight-infected potato leaves before (left) and after(right) MN Patch puncture. **(h)** Representative smartphone fluorescence images of the Samples. **(i)** The normalized fluorescence intensities of different reaction tubes.

Finally, DNA extracted via the microneedle patch was tested as the template using the smartphone-based detection device. DNA samples were collected by puncturing the lesion areas of *P. infestans*-infected potato leaves (**Fig. 7g**), while genomic DNA extracted from *P. infestans* mycelium by the CTAB method served as the positive control, and sterile water was used as the negative control. The fluorescence images of the reaction products at the endpoint are shown in **Fig. 7h**, and the corresponding normalized fluorescence intensities are presented in **Fig. 7i**, confirming that the smartphone-based detection device can successfully be applied for the detection of potato late blight.

## 4. Conclusions

In this study, we successfully developed a portable detection system integrating microneedle (MN) sampling with RPA-CRISPR-Cas12a technology. This is the first application of CRISPR-Cas12a for the detection of the oomycete pathogen *P. infestans*, filling a key gap in oomycete diagnostics. Compared with traditional molecular methods such as PCR and LAMP, the system offers several notable advantages:

Firstly, RPA and CRISPR-Cas12a reactions operate at 37 °C without requiring sophisticated thermal cyclers, simplifying equipment needs.

Secondly, the high specificity of CRISPR-Cas12a trans-cleavage ensures sensitive detection while reducing dependence on laboratory conditions.

Thirdly the MN patch shortens DNA extraction time from 3–4 hours (CTAB method) to 1 minute, with higher DNA extraction efficiency per milligram of leaf tissue.

Fourthly, the smartphone-based fluorescence device is low-cost (<800 RMB), lightweight (450 g), and enables automated "sample-in to result-out" detection in resource-limited field settings.

By integrating molecular diagnostics with portable hardware, we built a complete platform covering sampling, detection, and result analysis. Unlike conventional laboratory-dependent plant pathogen detection, this system provides a rapid field-deployable solution through efficient MN sampling, fast RPA-CRISPR-Cas12a reaction (within 90 minutes), and smartphone image analysis. The approach not only applies to potato late blight but also offers a replicable framework for detecting other pathogens such as fungi and bacteria.

In conclusion, this technology provides a critical tool for the early detection and control of potato late blight, enabling pathogen identification before visible symptoms develop and allowing timely intervention. Future work will aim to optimize microneedle DNA capture, enhance assay sensitivity, improve field stability, and expand applications to other crop pathogens, driving innovation in precision agriculture and plant protection.

## CRediT authorship contribution statement

**Jiangnan Zhao:** Methodology, Data curation, Investigation, Writing - original draft, and formal analysis. **Hanbo Xu:** Methodology, Investigation, Software, Writing - review & editing. **Cifu Xu:** Investigation, Software, Visualization, Validation. **Wenlong Yin:** Investigation, Software, Validation. **Laixin Luo:** Supervision. **Gang Liu:** Conceptualization, Project administration. **Yan Wang:** Conceptualization, Supervision, Project administration, Writing - review & editing.

## Declaration of competing interest

The authors declare that they have no known competing financial interests or personal relationships that could have appeared to influence the work reported in this paper.

## Acknowledgements

The authors sincerely thank funding support from the National Natural Science Foundation of China (Award # 62401569).